\documentclass[preprint]{ptephy_v2}%%%%%% to generate preprint number with ptep logo

\usepackage{gensymb}
\usepackage{graphics}  % Include figure files
\usepackage{dcolumn}  % Align table columns on decimal point
\usepackage{bm}  % bold math
\usepackage[mathlines]{lineno}  % Enable numbering of text and display math
\usepackage{upgreek}  % upright greek letters
\usepackage{multirow}  % multirow in table
\usepackage{xcolor}
\usepackage{placeins}  % \FloatBarrier
\usepackage{hyperref}
\usepackage{doi}

\begin{document}
%\linenumbers 
\title{Development of a Hermetic Gaseous Xenon Detector for Suppressing External Radon Background}

\author[1]{R.~Miyata}
\author[1]{K.~Fujikawa}
\author[1]{R.~Harata}
\author[1,2]{Y.~Itow\thanks{Present Address: Institute for Cosmic Ray Research, University of Tokyo, Kashiwa, Chiba, Japan}}
\author[1,2]{S.~Kazama}
\author[2]{M.~Kobayashi}

\affil[1]{Institute for Space-Earth Environmental Research, Nagoya University, Nagoya, Aichi, 464-8601, Japan }
\affil[2]{Kobayashi-Maskawa Institute for the Origin of Particles and the Universe, Nagoya University, Nagoya, Aichi, 464-8601, Japan \email{miyata.ryuta.b2@s.mail.nagoya-u.ac.jp, kobayashi.masatoshi.i9@f.mail.nagoya-u.ac.jp}}

\begin{abstract}
Radon-induced backgrounds, particularly from $^{222}$Rn and its beta-emitting progeny, present a critical challenge for next-generation liquid xenon (LXe) detectors aimed at probing dark matter down to the neutrino fog. 
To address this, we developed a compact hermetic gaseous xenon (GXe) detector. This device physically isolates the active volume from external radon sources by using a PTFE vessel sealed between two quartz flanges with mechanically compressed ePTFE gaskets.
To quantify radon sealing performance, we implemented a dual-loop GXe circulation system and conducted a 670-hour radon-injection measurement campaign. 
Radon ingress into the hermetic detector was monitored using electrostatic radon detectors and photomultiplier tubes (PMTs). 
From these two independent measurements, the steady-state ratios of the radon concentrations inside the hermetic detector to those outside were estimated to be \((1.1 \pm 0.1) \times 10^{-2}\) and \((1.1 \pm 0.2) \times 10^{-2}\), corresponding to radon-leakage flows of $(2.9 \pm 0.3)\times 10^{-11}$ and $(2.6 \pm 0.4)\times 10^{-11}$\,m$^{3}\,\mathrm{s}^{-1}$, respectively.
An extrapolation to a 60-tonne LXe TPC such as XLZD suggests that the radon leakage could amount to $1.2 \times 10^{-2}$\,mBq, which is negligible compared to the expected natural radon emanation inside the detector, typically 3\,mBq. 
These results demonstrate that flange-based mechanical sealing provides an effective solution for realizing radon-isolated inner detectors in large-scale LXe experiments.
\end{abstract}

\subjectindex{dark matter, liquid xenon, low background}

\maketitle

\section{Introduction}
Dark matter, an elusive constituent of the universe, is thought to account for approximately 85\,\% of its total matter content. Despite compelling indirect evidence from astrophysical observations, its particle nature remains unknown. One of the most promising dark matter candidates is the Weakly Interacting Massive Particle (WIMP), predicted by many extensions of the Standard Model~\cite{wimp_model}. Direct detection experiments aim to observe rare WIMP--nucleus scattering events, requiring detectors with extremely low background levels.

Dual-phase liquid xenon (LXe) time projection chambers (TPCs) have become the leading technology in this field due to their scalability, self-shielding capabilities, and excellent discrimination between nuclear and electronic recoils~\cite{RevModPhys.82.2053}. By detecting prompt scintillation light (S1) and delayed electroluminescence signals (S2) produced by ionization electrons extracted into the gas phase, these detectors enable three-dimensional reconstruction of interaction positions, which facilitates fiducialization of the active volume and effective suppression of external radioactive backgrounds.

As the target mass of LXe detectors increases toward the multi-tonne scale, mitigating radioactive backgrounds becomes increasingly important. In current-generation LXe TPCs such as XENONnT~\cite{XENON:2023cxc}, PandaX-4T~\cite{PandaX:2024qfu}, and LZ~\cite{LZ:2022dm}, one of the dominant backgrounds in the WIMP search originates from the beta decays of $^{214}$Pb, a daughter of $^{222}$Rn. The $^{222}$Rn isotope, with a half-life of 3.82 days, emanates continuously from detector materials containing trace amounts of $^{226}$Ra and mixes uniformly into the LXe target. While existing experiments have achieved impressive control over internal radon levels, down to 0.9\,$\mu$Bq/kg~\cite{xenon_radon}, future observatories such as PandaX-xT~\cite{PandaXT}, DARWIN~\cite{Aalbers:2016jon}, and XLZD~\cite{XLZD:design}, with target masses exceeding 40\,tonnes, aim to reach sensitivities at or below the so-called neutrino fog~\cite{PhysRevLett.127.251802}. At that level, interactions from astrophysical neutrinos become the dominant irreducible background. To ensure that radon-induced backgrounds remain subdominant, the radon concentration must be reduced to below 0.1\,$\mu$Bq/kg, an order of magnitude improvement over current benchmarks.

To achieve the stringent radon background requirements of next-generation LXe detectors, a combination of mitigation techniques is considered essential. These include material selection~\cite{XENON:screening, LZ:screening, nEXO_material}, surface treatment~\cite{surface_coating}, and active removal systems such as cryogenic distillation~\cite{xenon_radon, lowrad}. As part of this multi-pronged approach, geometry-based methods that physically limit radon transport into the active volume have attracted growing attention.

One such concept is the hermetic detector, where the active LXe target is enclosed within a sealed inner chamber, physically isolated from surrounding components that may emit radon. This reduces the effective contact area between the active xenon and potential sources of contamination. A pioneering implementation by the Freiburg group employed cryofitting techniques, exploiting differential thermal contraction between PTFE and quartz or metal components to achieve press-fit seals~\cite{freiburg_hermetic}. Their prototype demonstrated effective radon suppression, and scaling studies suggested feasibility for large detectors such as XLZD. However, the performance of such systems is critically dependent on minimizing the leakage flow, the volume exchange rate between inner and outer volumes.

In this work, we present an alternative implementation of the hermetic detector concept using conventional sealing technology, based on bolted flanges and expanded PTFE (ePTFE) gaskets. We constructed a small-scale prototype to evaluate its radon sealing performance in a gaseous xenon (GXe) environment at room temperature and to assess its potential for integration into next-generation ultra-low-background LXe detectors such as XLZD.

The structure of the paper is as follows. Section~\ref{sec:detector_design} describes the detector design and sealing strategy. Section~\ref{sec:setup} outlines the experimental setup and data acquisition system. Section~\ref{sec:analysis} details the modeling of radon leakage, the analysis procedure, and the results of radon sealing performance of the hermetic detector. 
In Sec.~\ref{sec:discussion} we discuss the implications for scaling to $\mathcal{O}(10)$-tonne detectors, and conclude in Sec.~\ref{sec:conclusion} with a summary and an outlook on future applications.

%%%%%%
\section{Design of the Hermetic Detector}
\label{sec:detector_design}

In conventional dual-phase LXe TPCs, the active target volume is in direct contact with many components, including electrodes, cables, reflectors, and cryostat structures that can emit $^{222}$Rn. To suppress radon-induced backgrounds beyond the limits of material purification and online removal techniques, we propose a detector geometry where the sensitive LXe region is housed inside a hermetically sealed inner chamber. In this design, only a carefully selected subset of low-radon materials (e.g., PTFE and quartz)~\cite{XENON:screening,xmass_pmt} is in direct contact with the active xenon. 
The rest of the detector components, which may emit significant radon, are placed in an outer volume physically isolated from the target. Our concept is based on a geometric barrier approach: instead of reducing the radon emanation rate from every component, we reduce the effective area through which radon can enter the active region. A well-sealed boundary between the inner and outer xenon volumes acts to suppress the inward transport of radon via diffusion or convection. In such a configuration, the dominant parameter governing performance is the leakage flow, defined as the volume exchange rate of xenon across the sealed interface. 

\begin{figure}[ht!]
        \centering\includegraphics[width=0.8\linewidth]{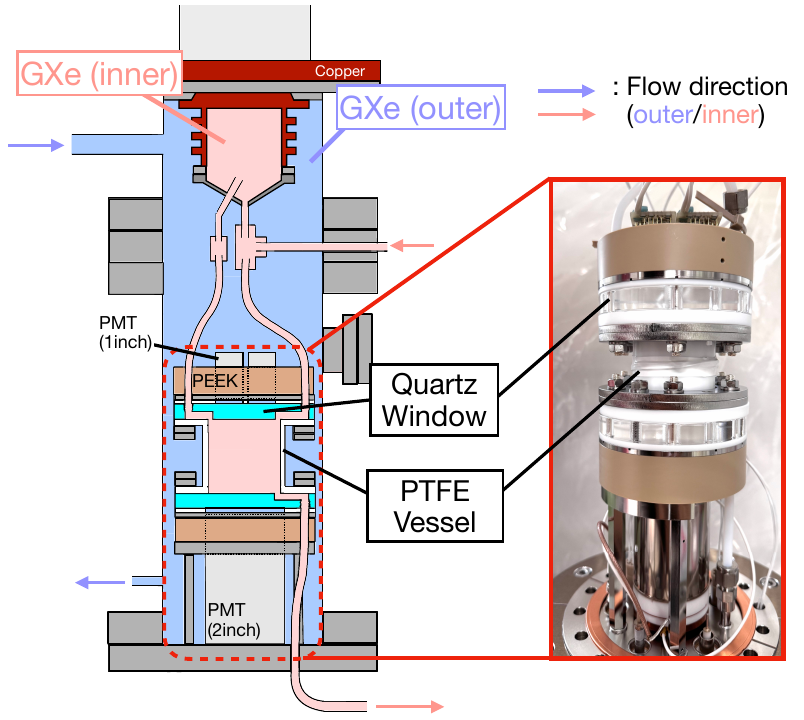}
        \caption{Schematic diagram (left) and photograph (right) of the hermetic detector developed in this study.}
        \label{fig:det_picture}
\end{figure}

While previous studies by the Freiburg group have explored this concept using cryofit seals based on thermal contraction~\cite{freiburg_hermetic}, we propose a different implementation. Our design employs conventional sealing technology, bolted flanges with compressible ePTFE gaskets, to construct a hermetic detector with robust sealing performance under both room-temperature and cryogenic conditions. To evaluate this design, we constructed a small-scale prototype hermetic detector. The detector has an internal diameter of 50\,mm and a height of 63\,mm, corresponding to a gas volume of approximately 0.1\,L. 
Figure~\ref{fig:det_picture} shows a schematic diagram and a photograph of the hermetic detector developed in this study.

The detector vessel is made of PTFE and is sealed at the top and bottom by two VUV-transparent quartz flanges, each 12.7\,mm thick. These windows allow the transmission of xenon scintillation light to photodetectors positioned outside the sealed volume. The flanges are mechanically fastened to the PTFE vessel using stainless-steel bolts and washers, compressing an ePTFE gasket between polished surfaces to create a mechanical seal. PTFE and quartz were selected for their radiopurity~\cite{XENON:screening,xmass_pmt}, chemical inertness, and low outgassing properties, which make them ideal materials for use in noble-gas environments. Both materials are widely used in modern LXe detectors, such as XENONnT, where PTFE functions as the inner reflector and quartz serves as the photomultiplier tube (PMT) window material.
In our previous study~\cite{sato_hermetic}, we demonstrated that an LXe TPC incorporating a quartz vessel with flange-like plates, though lacking any dedicated sealing mechanism, functioned with no measurable degradation in detector performance due to the presence of quartz.

Special attention was given to optimizing the sealing configuration. The ePTFE gasket was compressed in a flat-on-flat geometry using six symmetrically placed M6 bolts. To evaluate the leak performance, helium leak tests were conducted using a helium leak detector (Pfeiffer Adixen ASM 310). Instead of directly using helium spray, the tests were performed using the natural helium present in ambient air. During the test, the detector was pumped and maintained at approximately 1\,Pa. To provide a reference for comparison, the baseline helium leak rate, originating from atmospheric helium permeation, was measured using the leak detector. This baseline provided good reproducibility and enabled reliable relative comparisons between different sealing configurations.

\begin{figure}[ht!]
  \centering
    \centering
    \includegraphics[width=0.6\linewidth]{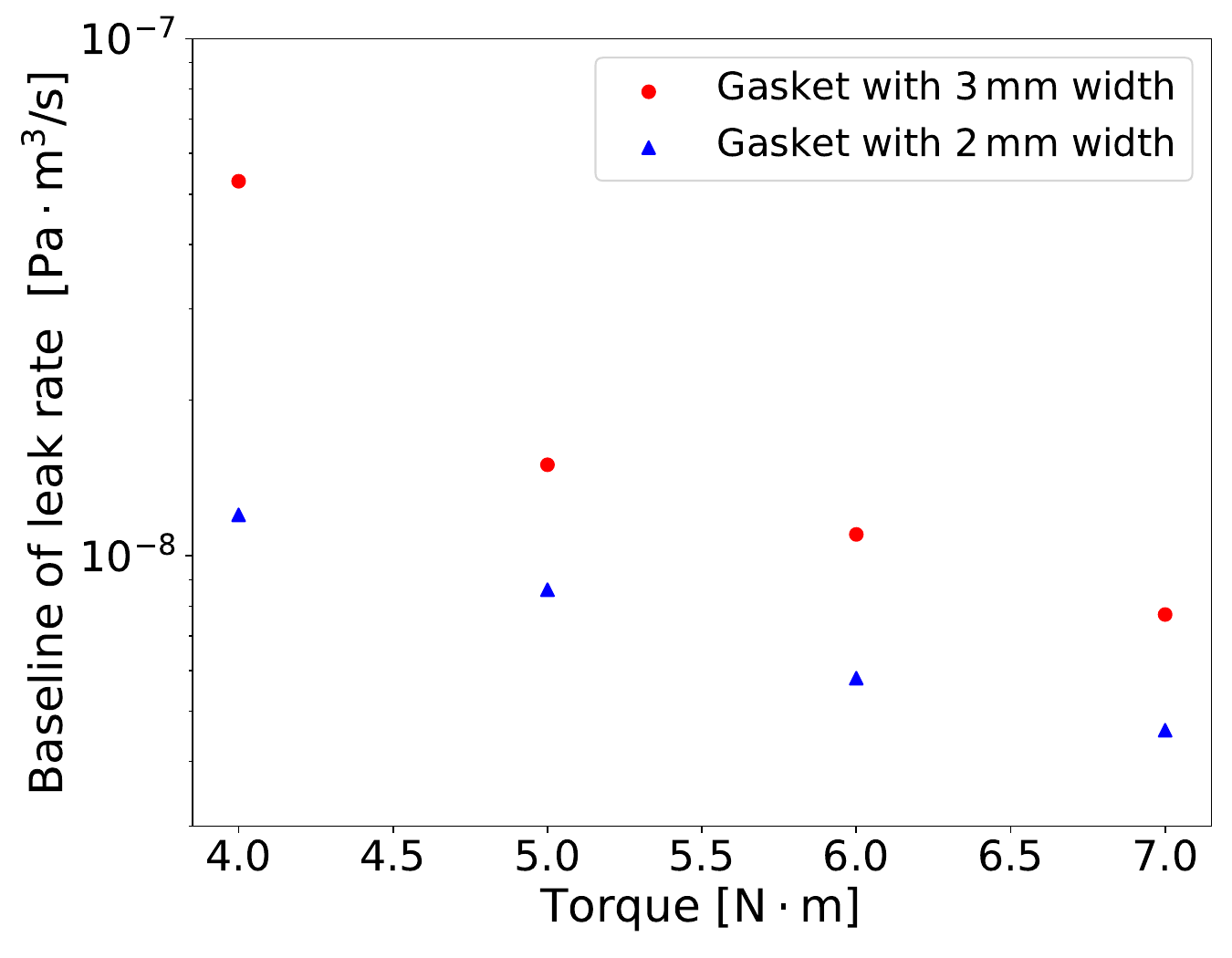}
    \caption{Baseline of helium leak rate as a function of the torque applied to the bolts. Red and blue points correspond to gasket widths of 3\,mm and 2\,mm, respectively.}
   \label{fig:leakrates_gasket}
\end{figure}

We investigated the dependence of the baseline leak rate on the torque applied to the bolts and the gasket width. 
Gaskets with widths of 2\,mm (Gore Hyper-Sheet Gasket) and 3\,mm (Valqua 7GP66S) were tested. 
The flange was tightened with bolts torqued to 4.0, 5.0, 6.0, and 7.0\,N$\cdot$m. 
Figure~\ref{fig:leakrates_gasket} shows the results of the test as a function of the applied torque. 
The narrower gasket with a width of 2\,mm consistently exhibited better sealing performance compared to the 3\,mm gasket.

Because a crack was observed in the flange at 7.5\,N$\cdot$m, the test was stopped at 7.0\,N$\cdot$m. 
A configuration with a 2\,mm gasket and a torque of 5.0\,N$\cdot$m was selected for the hermetic detector to mitigate the risk of mechanical failure. 
The pressure on the gasket in this configuration was calculated to be 59\,MPa, which satisfies the recommended minimum pressure of 30\,MPa specified by the gasket manufacturer.

To assess the thermal stability, cooling tests were conducted. 
The detector was cooled to $-100\,^\circ$C using liquid nitrogen at a rate of 23\,$^\circ$C per hour. 
After warming back to room temperature, no measurable degradation in sealing performance was observed. 
In addition, the helium leak rate was monitored weekly over a one-month period to evaluate the long-term stability. 
The leak rate remained stable within 18\,\% during the monitoring period, demonstrating that the sealing configuration is suitable for long-term operation.

\section{Experimental Setup and Data Acquisition}
\label{sec:setup}
\subsection{Gas circulation system}
\label{sec:circulation_system}
\begin{figure}[ht!]
        \centering
        \includegraphics[width=0.75\linewidth]{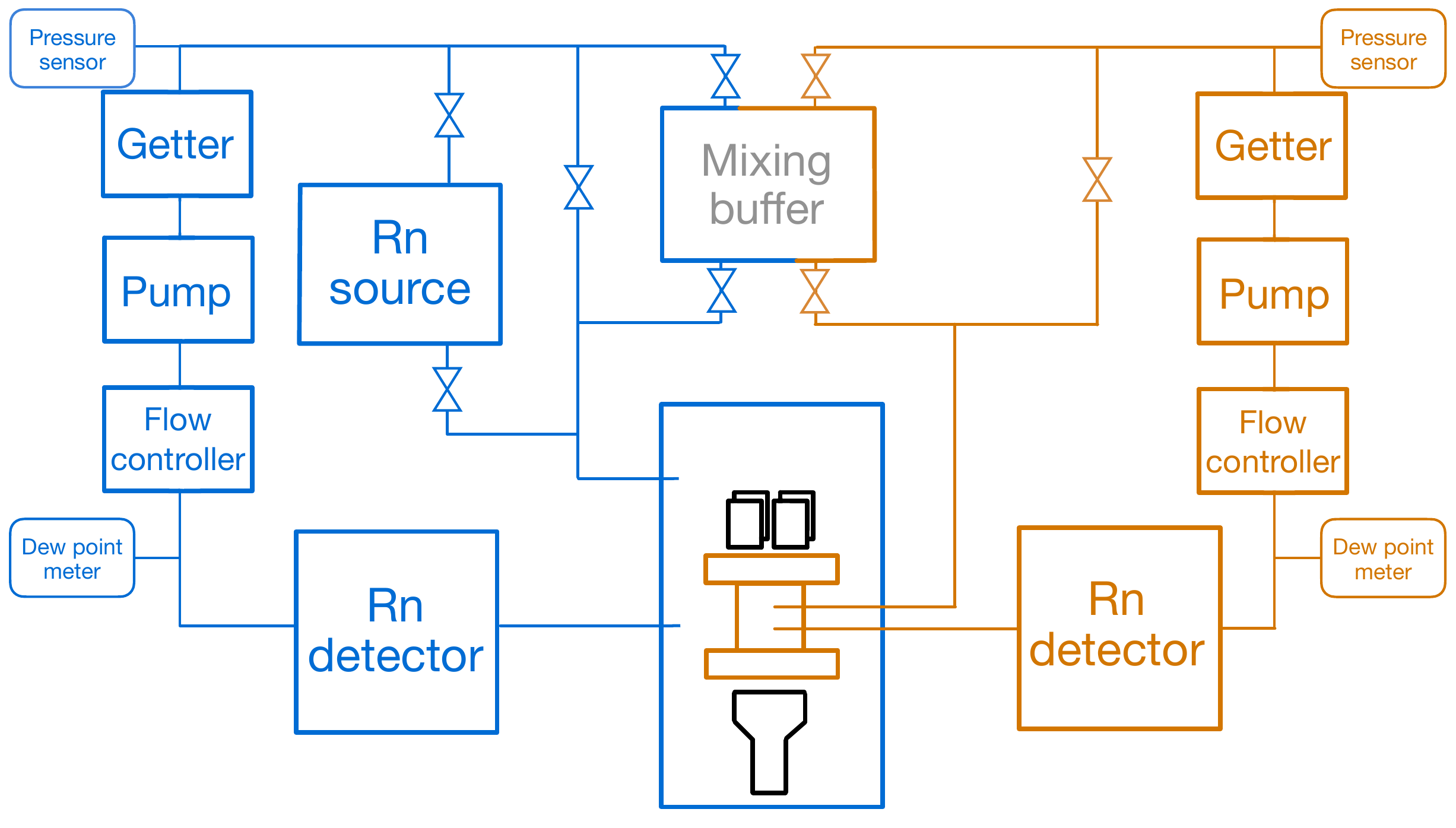}
        \caption{Schematic overview of the gas circulation system. The brown and blue lines correspond to the inner and outer circulation loops, respectively. The mixing buffer is connected to both loops and allows for controlled exchange between them.}
        \label{fig:circulation}
\end{figure}

To evaluate the radon sealing performance of the detector described in Sec.~\ref{sec:detector_design}, 
a dedicated gas-handling system was constructed and integrated with the hermetic detector. 
A schematic diagram of the circulation system is shown in Fig.~\ref{fig:circulation}. 
The system was designed to independently circulate and purify xenon, monitor its pressure and dew point, and allow simultaneous, quantitative measurements of radon concentrations in both volumes.

The gas circulation system consists of two independent loops: one for the inner volume (the gas enclosed within the hermetic detector) and one for the outer volume (the gas surrounding the hermetic detector). The use of separate circulation lines ensures that no direct gas exchange occurs between the inner and outer volumes during operation. Each loop includes a metal-bellows pump (IBS MB-151AL or MB-111) and a zirconium-based hot-metal getter (NuPure Omni 1000 or 600) for continuous circulation and purification of the GXe. To monitor and control the system, each loop is equipped with a high-precision pressure sensor (Pureron PC-304A), a dew point transmitter (Michell PURA or Tekune TK-100), and a mass-flow controller (Lintec MC-730MC).  

Although the test described in this study was performed using GXe at room temperature, the overall experimental setup is also compatible with future tests under LXe conditions. 
Each circulation loop is connected to its respective detector volume via a dedicated heat exchanger (Hisaka Works BXC-024-NU-20).
In addition, the entire detector assembly is enclosed within a large vacuum chamber for thermal insulation and is designed to be coupled to a pulse-tube refrigerator in future studies, allowing the xenon to be cooled to its liquid phase at approximately 175\,K.

To introduce radon into the system, the radon source was installed in the outer loop. The source consists of ceramic beads impregnated with \(^{226}\)Ra, providing a stable radon emanation rate of approximately 10\,Bq contained in the stainless steel chamber. GXe mixed with radon emanating from the source was circulated through the outer loop, establishing a uniform radon concentration in the outer volume surrounding the hermetic detector. Several interconnection valves and the mixing buffer between the inner and outer volumes are included in the circulation system to enable uniform injection of radon into the entire system for calibration purposes. A fine-particle filter (Pureron UCSX-M) with a mesh size of 3\,nm is placed both upstream and downstream of the source to prevent the transfer of radioactive particulates into the main system.

\subsection{Radon measurement systems and data acquisition}
\label{sec:radon_detector_DAQ}
The radon concentration in the detector was estimated using an electrostatic collection-type radon detector developed by Super-Kamiokande group~\cite{Rn_SK}. Since 90\,\% of radon daughters such as \(^{214}\)Po are produced as positive ions~\cite{radon_daughters}, they can be collected by an electric field onto a PIN photodiode, where their alpha decays are detected directly. In contrast, \(^{222}\)Rn itself is a neutral noble gas atom and therefore cannot be collected by the electrostatic detector in the same manner. The detector has a volume of 1\,L and operates at a collection voltage of 112\,V. Its nominal detection efficiency is approximately 10\,\%. Since the efficiency degrades in the presence of water vapor due to ion neutralization, the humidity of the system must be kept sufficiently low. 
The dew point was monitored and kept below $-80\,^{\circ}$C throughout the experiment, ensuring that there was no significant degradation in the performance of the radon detectors.
Two radon detectors were installed independently, one in each loop. The charges generated in the PIN photodiode are first amplified and shaped by a preamplifier~\cite{Rn_SK}, and then digitized by a CAEN V1724 module operating at a 100\,MHz sampling rate.

In addition to the radon detectors, PMTs are installed above and below the hermetic detector to detect GXe scintillation light produced by alpha decays of radon and its progeny. Four 1-inch PMTs (Hamamatsu R8520) are mounted on the top of the detector, while one 2-inch PMT (Hamamatsu R10789) is installed on the bottom. Both types of PMTs were developed for and have been widely used in LXe experiments, offering a high quantum efficiency (approximately 30\,\%) for xenon scintillation light. Typical operating voltages are 780\,V for the top PMTs and 1160\,V for the bottom PMT, corresponding to a gain of $10^{6}$ for single photoelectrons. The PMT outputs are also digitized using a CAEN V1724 digitizer operating at a sampling rate of 100\,MHz. These PMTs complement the radon detectors by providing an independent measurement of alpha activity inside and outside the hermetic detector.

Data acquisition for the radon detectors and PMTs is triggered by a logical OR of the signals from the two radon detectors and the bottom PMT. Given the high energy of alpha particles, only the bottom PMT is used for triggering, while the top PMTs are excluded. Although the top PMTs are not included in the trigger logic, their signals are recorded and used in the offline reconstruction of alpha decay events.

\subsection{Operation}
\label{sec:operation}
The operation was divided into three distinct phases: a background run, a radon run, and a calibration run.
In the first phase (background run), the radon concentration in the system was estimated without any radon source. 
The purpose of this run was to confirm that the natural emanation of detector components such as piping, circulation pumps, and other materials was sufficiently low. 
Prior to the measurement, both circulation loops were operated through their respective getters for several hours to reduce impurities. 
The measurement was conducted over a period of 42 hours.

\begin{figure}[ht!]
  \centering
    \centering
    \includegraphics[width=0.7\linewidth]{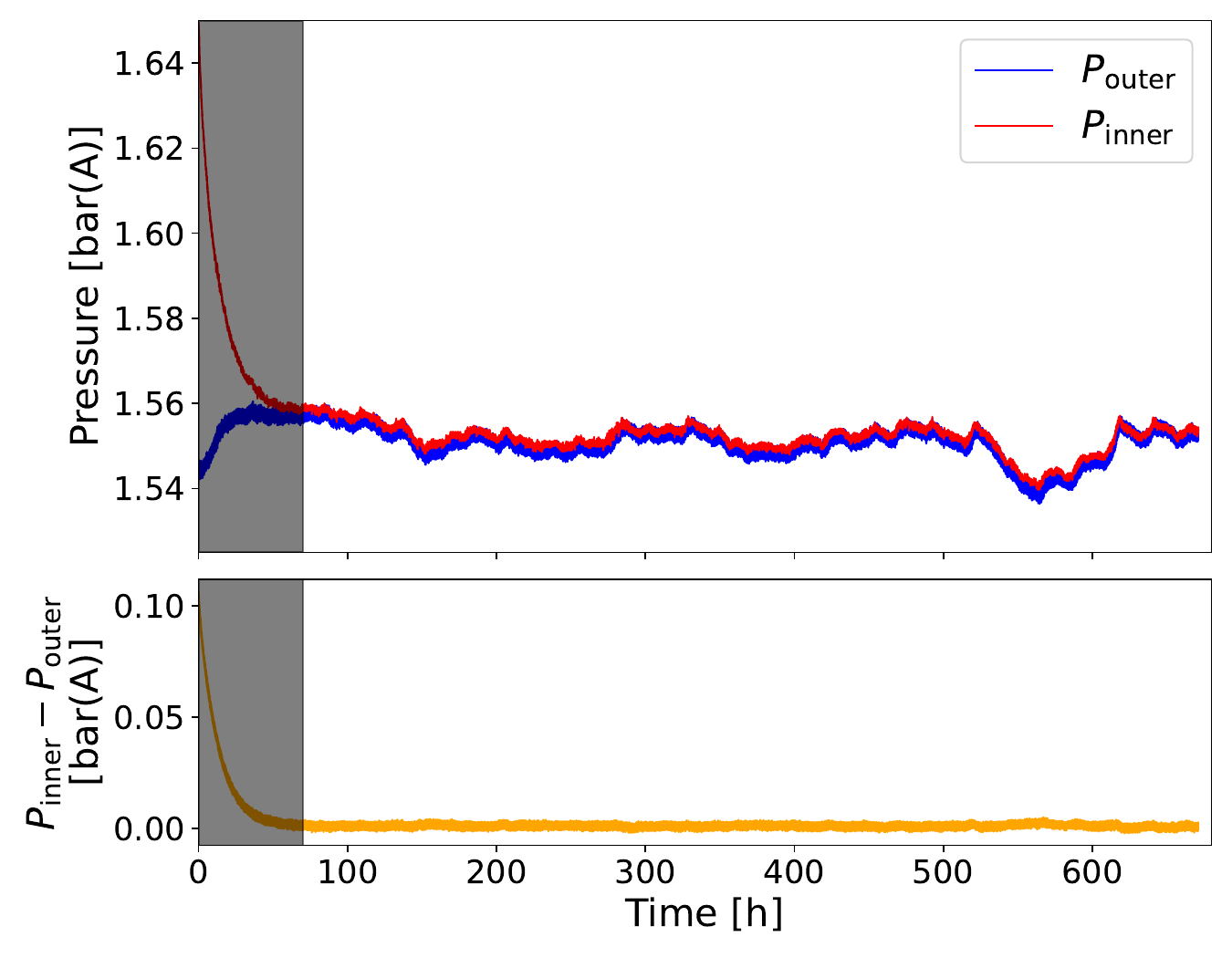}
    \caption{Pressure of the system during the radon run. Upper: Red and blue lines represent the pressure for inner volume ($P_{\mathrm{inner}}$) and outer volume ($P_{\mathrm{outer}}$), respectively. Lower: Pressure difference between inner and outer volume. Data within the shaded region were excluded from the analysis. See the text for details.}
   \label{fig:pressure}
\end{figure}

In the second phase (radon run), the radon source was connected to the outer loop, allowing radon to diffuse into the system. 
Figure~\ref{fig:pressure} shows the pressure of the inner and outer volumes during this run. 
The pressure of the outer volume dropped by approximately 0.1\,bar immediately after opening the source, due to the small pressure difference between the system and the source chamber. 
As described in Sec.~\ref{sec:detector_design}, the hermetic detector we developed is not perfectly vacuum-tight; accordingly, the pressure difference was gradually relieved.
In the shaded region of Fig.~\ref{fig:pressure}, where the pressures inside and outside the hermetic detector differ, a pressure-driven flow of GXe from inside the detector to the outside can occur, potentially affecting the measurement of diffusion-driven radon leakage.
Therefore, to ensure full stabilization, only data taken after 70\,hours were used for the analysis. 
Radon concentrations were continuously monitored for 670\,hours, corresponding to approximately five radon lifetimes.

In the final phase (calibration run), the interconnecting valves and the mixing buffer between the inner and outer volumes were opened for five days. 
During this run, the radon concentrations inside and outside the hermetic detector were equalized. 
This enabled us to calibrate for the differences in measurement conditions between the inner and outer volumes, such as the detection efficiencies of the radon detectors and the sensitive volumes for measurements with the PMTs. 
Further details of the analysis are presented in Sec.~\ref{sec:analysis}.

\section{Analysis and Results}
\label{sec:analysis}
In this section, we present analyses to quantify the radon sealing performance and their results. 
As described in Sec.~\ref{sec:detector_design}, we use the leakage flow $f$ as a quantitative metric of the radon sealing performance, with units of m$^3\,\mathrm{s}^{-1}$.
The number of radon atoms that flow per unit time from the outside to the inside can be expressed in terms of $f$ as follows:
\begin{equation}
    f\cdot \left ( \frac{N_\mathrm{O}}{V_\mathrm{O}}-\frac{N_\mathrm{I}}{V_\mathrm{I}}\right),
\end{equation}
where \( N_i \) is the number of radon atoms in volume \( V_i \); therefore, \( N_i \)/\( V_i \) represents the radon concentration. The subscripts \(\mathrm{O}\) and \(\mathrm{I}\) refer to the outside and inside of the hermetic detector, respectively.

The time evolution of $N_\mathrm{I}$ is governed by two competing processes: the ingress of radon atoms through microscopic leaks and the radioactive decay of radon already present in the inner volume. 
Consequently, the time evolution of \( N_{\mathrm{I}} \) is described by the differential equation:
\begin{equation}
\frac{dN_{\mathrm{I}}(t)}{dt} = f\cdot\left( \frac{N_{\mathrm{O}}(t)}{V_{\mathrm{O}}} - \frac{N_{\mathrm{I}}(t)}{V_{\mathrm{I}}} \right)-\frac{1}{\tau} N_{\mathrm{I}}(t),
\end{equation}
where \( \tau = 132 \)\,hours is the lifetime of \(^{222}\)Rn. 
Assuming that the time dependence of \( N_{\mathrm{O}} \) is negligible due to constant radon emanation from the source,
the ratio of radon concentrations inside to outside the hermetic detector, \( R(t) \), can be expressed as
\begin{equation}
R(t) \equiv \frac{\frac{N_{\mathrm{I}}(t)}{V_{\mathrm{I}}}}{\frac{N_{\mathrm{O}}}{V_{\mathrm{O}}}} = \frac{\tau f}{V_{\mathrm{I}} + \tau f}\left(1 - e^{-\frac{V_{\mathrm{I}}+\tau f}{\tau V_{\mathrm{I}}}t}\right)+R(0).
\label{eq:eq_fit}
\end{equation}
This expression describes the exponential approach of the ratio of radon concentrations to its steady-state value as 
\begin{equation}
    \lim_{t \to \infty} R(t) = \frac{\tau f}{V_{\mathrm{I}} + \tau f}+R(0).
\end{equation}
Therefore, $f$ can be estimated by observing the time evolution of the ratio of radon concentrations. 

In the following section, we will introduce each step of the analysis to evaluate the radon sealing performance.
First, Sec.~\ref{sec:analysis_alpha} describes the selection criteria of alpha decay events for the data acquired with the electrostatic radon detectors and with the PMTs. Second, Sec.~\ref{sec:conversion} explains how to estimate the ratio of radon concentrations from the observed count rates of alpha decay. Finally, Sec.~\ref{sec:time_evolution} presents the measured time evolution of the ratio of radon concentrations and the results on the sealing performance of radon.

\subsection{Selection of alpha events}
\label{sec:analysis_alpha}
\subsubsection{Analysis of electrostatic radon detector data}
\label{analysis_radondet}
\begin{figure}[ht!]
  \centering
    \centering
    \includegraphics[width=0.99\linewidth]{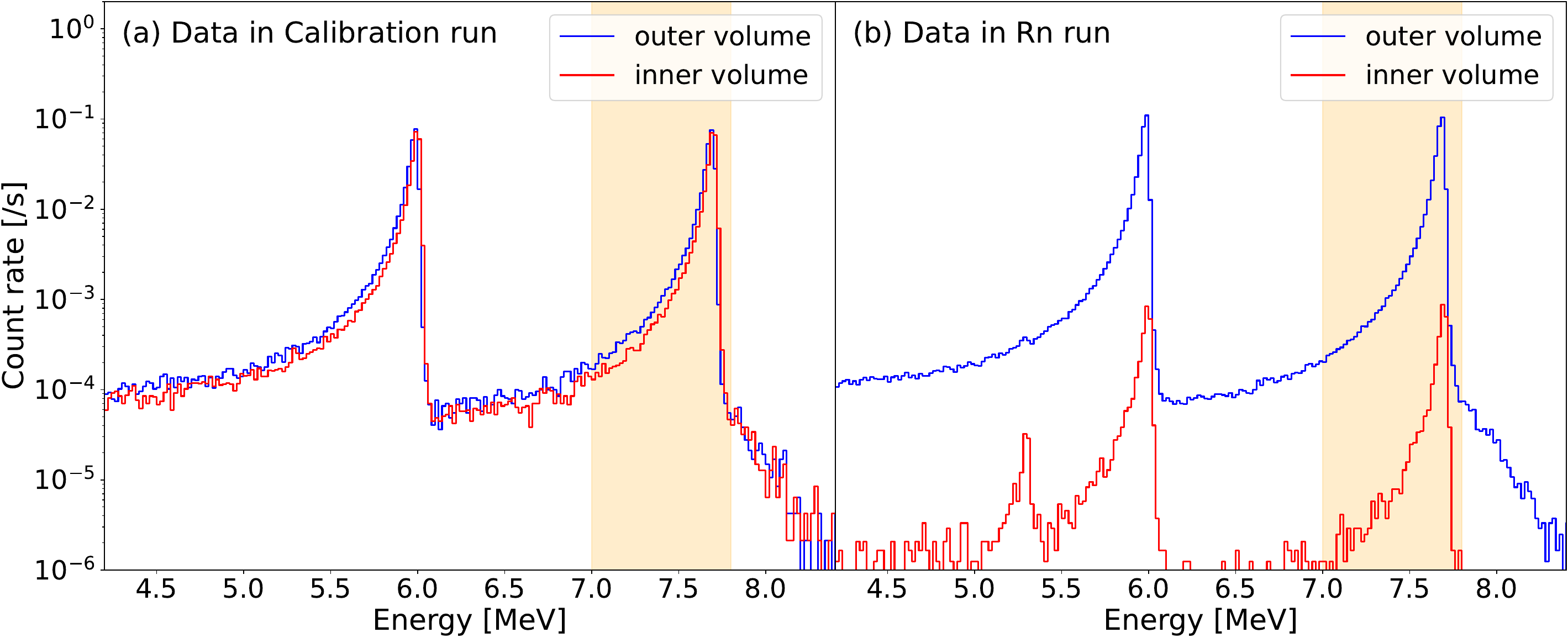}
    \caption{Energy spectra of alpha particles observed by the electrostatic radon detectors during (a) the calibration run and (b) the radon run. Blue and red histograms correspond to the outer and inner volumes, respectively. The shaded regions indicate the energy window used in the analysis, corresponding to the range [7.0,\,7.8]\,MeV.}
   \label{fig:spectrum_radon_detector}
\end{figure}

As described in Sec.~\ref{sec:radon_detector_DAQ}, the electrostatic radon detector is sensitive to the alpha decays of $^{218}$Po and $^{214}$Po, with energies of 6.0 and 7.7\,MeV, respectively.
The analysis follows the method described in Ref.~\cite{Rn_SK}. 
Note that only the $^{214}$Po peak was used for this evaluation, since the $^{218}$Po peak can be contaminated by the alpha decays of $^{210}$Po, which emits 5.3\,MeV alphas and overlaps in energy. 
Figure~\ref{fig:spectrum_radon_detector} shows the energy spectra of alpha particles observed by the radon detectors during the calibration and radon runs, along with the shaded region used to evaluate the ratio of radon concentrations. 
The number of $^{214}$Po events was obtained by integrating the counts in the range [7.0,\,7.8]\,MeV.

\subsubsection{Analysis of PMT data}
\label{analysis_pmt}
\begin{figure}[ht!]
   \includegraphics[width=0.98\columnwidth]{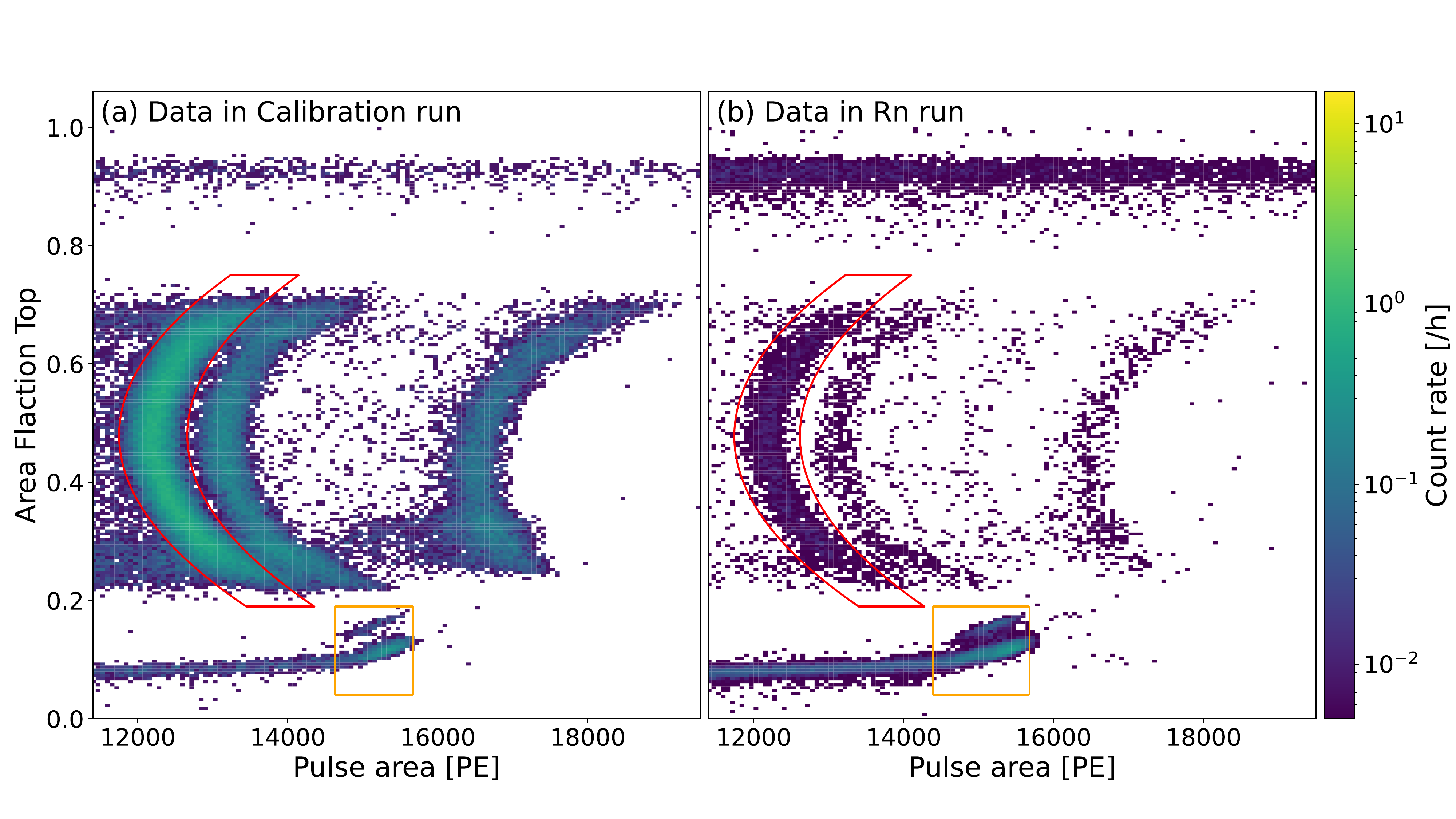}
    \caption{Distributions of pulse area and area fraction top (AFT) for (a) the calibration run and (b) the radon run. The red and orange regions indicate the areas used to estimate the radon activity in the inner and outer volumes, respectively. See the text for details.}
\label{fig:aft_vs_pe}
\end{figure}

As described in Sec.~\ref{sec:radon_detector_DAQ}, alpha particles originating from radon and its progeny can be detected through the scintillation light of GXe with PMTs.
The PMTs detected events occurring inside the hermetic detector, as well as those originating in the gap between the quartz window and the PMTs.
To distinguish between events occurring inside and outside the hermetic detector, the Area Fraction Top (AFT), defined as the fraction of detected photons observed by the top PMTs relative to the total, was used in this analysis. 
Because of the difference in light-collection acceptance between the top and bottom PMTs, events occurring closer to the top PMTs yield larger AFT values.
Figure~\ref{fig:aft_vs_pe} shows the distribution of pulse area and AFT in the calibration and radon runs.
Three distinct populations of AFT can be identified: AFT $<$ 0.19, 0.19 $\leq$ AFT $\leq$ 0.75, and 0.75 $<$ AFT, corresponding to events occurring in the gap between the bottom PMT and the hermetic detector, inside the hermetic detector, and in the gap between the top PMTs and the hermetic detector, respectively.

In this analysis, the Cherenkov light in the detector’s quartz window produced by cosmic-ray muons can be background events.
Pulse-shape discrimination was performed based on the pulse width, defined as the time interval during which the pulse amplitude exceeds 10\,\% of its maximum. 
Signals with characteristically short widths were rejected, because the time constant of the Cherenkov process is much shorter than that of GXe scintillation.
This procedure effectively reduced the muon-induced background to a negligible level.

The radon concentration in the inner volume was determined from events within the red region shown in Fig.~\ref{fig:aft_vs_pe}, attributed to $^{222}$Rn alpha decays. Two neighboring bands correspond to polonium decays.
Although alpha-particles produced in each decay are mono-energetic, the observed pulse-area within the hermetic detector varies with AFT and reaches a minimum around AFT $\sim$\,0.5. 

The radon concentration in the outer volume was determined from events within the orange region shown in Fig.~\ref{fig:aft_vs_pe}. 
The gap between the bottom PMT and the hermetic detector is approximately 3.5\,mm in height and 50\,mm in diameter.
Since the stopping range of alpha particles in GXe is $\sim$\,14\,mm, only those traveling horizontally form a distinct peak. Alpha particles moving in other directions strike detector components before depositing their full energy, producing a low-energy tail outside the selection box.
Furthermore, because the bottom PMT is positioned very close to this gap region, its acceptance for scintillation photons generated there is significantly enhanced, often leading to signal saturation.  
As a result, the pulse-area distributions of the three alpha decays of \(^{222}\)Rn, \(^{214}\)Po, and \(^{218}\)Po overlap and cannot be resolved in the orange region. 
Although there is a similar gap between the top PMTs and the hermetic detector, it is divided into four sections to accommodate individual 1-inch PMTs. 
Alpha particles originating in this upper gap are therefore even more likely to strike detector components before depositing their full energy, and no distinct peak is formed; therefore, events from this region were excluded from the analysis.

To estimate radon-leakage flow using Eq.~\ref{eq:eq_fit}, we evaluated the ratio of radon concentrations from the event rates observed in the red and orange regions of Fig.~\ref{fig:aft_vs_pe}. 
Due to the different characteristics of alpha events in the inner and outer volumes as explained above, the observed event rates are expected to differ even with the same radon concentration. 
To correct for this difference, we used calibration data in which the radon concentrations in both volumes were equalized. 
A detailed discussion of this correction is presented in the next section.

\subsection{Estimation of the ratio of radon concentrations}
\label{sec:conversion}
As described in the previous sections, evaluating the leakage flow requires estimating the ratio of radon concentrations while accounting for the difference in detector responses between the inner and outer volumes. 
In this section, we describe how to estimate this ratio using the event rates of alpha decays inside and outside the hermetic detector.
Because the radon concentration is proportional to the count rate of alpha decay, it can be expressed as follows:
\begin{equation}
    \frac{N_{i}}{V_{i}}=A_{i}C_{i},
\label{eq:proportional}
\end{equation}
where ${C}$ is the count rate of alpha events obtained by electrostatic radon detectors or PMTs, and ${A}$ is defined as a proportionality constant. 
Therefore, the ratio of radon concentrations can be written as follows:
\begin{equation}
    R(t)\equiv\frac{\frac{N_\mathrm{I}(t)}{V_\mathrm{I}}}{\frac{N_\mathrm{O}(t)}{V_\mathrm{O}}}=\frac{A_\mathrm{I}}{A_\mathrm{O}}\frac{C_\mathrm{I}(t)}{C_\mathrm{O}(t)}\equiv r\frac{C_\mathrm{I}(t)}{C_\mathrm{O}(t)},
\label{eq:count_rate}
\end{equation}
where $r$ is the ratio of proportionality constants, and needs to be estimated separately for the measurement with radon detectors and with PMTs. 
For the electrostatic detectors, $r$ reflects the small difference in detection efficiency between the two detectors, and is expected to be close to unity.  
For the PMT measurements, $r$ represents the differences in sensitive volumes and in the characteristics of alpha-event observation inside and outside the hermetic detector as explained in Sec.~\ref{analysis_pmt}.

This ratio of proportionality constant $r$ can be obtained using data from the calibration run, in which the radon concentrations inside and outside the hermetic detector were equalized so that $\frac{N_\mathrm{I}}{V_\mathrm{I}} = \frac{N_\mathrm{O}}{V_\mathrm{O}}$ is satisfied. 
Therefore, $r$ can be expressed as
\begin{equation}
    r = \frac{C^{\prime}_\mathrm{O}}{C^{\prime}_\mathrm{I}},
\label{eq:proportionality_constant}
\end{equation}
where $C^{\prime}$ denotes the event rate of alpha decays during the calibration run.
The obtained values of r were $1.071\pm 0.005$ and $(2.41\pm 0.01)\times 10^{-2}$ for measurement with radon detectors and PMTs, respectively. 

\subsection{Results on the radon sealing performance}
\label{sec:time_evolution}
By fitting $R(t)$ obtained in Sec.~\ref{sec:conversion} with Eq.~\ref{eq:eq_fit}, $f$ can be estimated as a direct measure of the hermeticity of the detector. 
The fits for the two datasets, one obtained with electrostatic radon detectors and the other with PMTs, were performed independently with $\chi^2$ minimization method. Each dataset was binned into 10-hour intervals, and only data after the first 70 hours, by which the detector condition had stabilized (see Sec.~\ref{sec:operation}), were included in the fit.
The contribution of natural radon emanation was negligible for this analysis, given that the observed alpha rate during the background run was less than 5\,\% of that during the radon run.

\begin{figure}[ht!]
  \centering
  \includegraphics[width=\columnwidth]{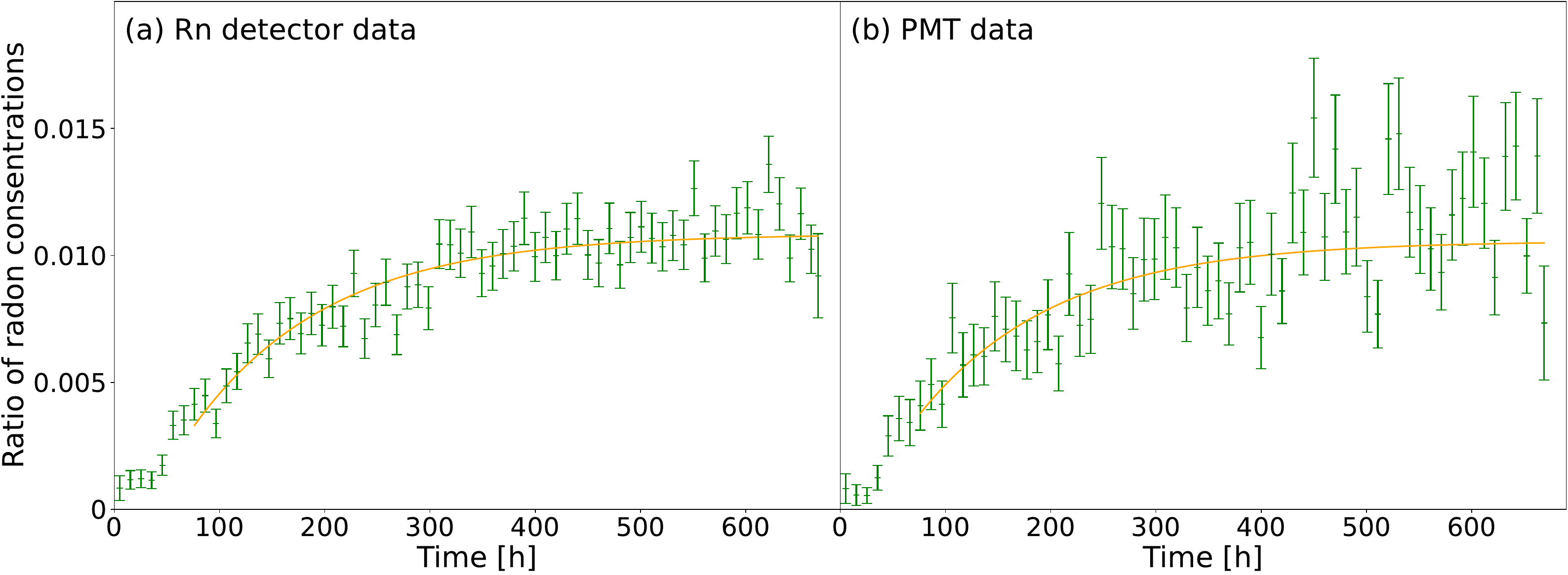}
  \caption{Time evolution of the ratio of radon concentrations in inner and outer volumes. Green points show the observed ratio, and the orange curve shows the best-fit result derived from Eq.~\ref{eq:eq_fit}. (a): Result obtained using electrostatic radon detectors, (b): Result obtained using PMTs. See the text for more details.}
  \label{fig:time_evolution}
\end{figure}

Figure~\ref{fig:time_evolution}\,(a) and (b) show the time evolution of the ratio of radon concentrations measured with electrostatic radon detectors and PMTs, respectively, together with the best-fit curves. 
The best-fit values of \( f \) were obtained as \((2.9 \pm 0.3) \times 10^{-11}\) and \((2.6 \pm 0.4) \times 10^{-11}\) \,$\mathrm{m}^{3}\,\mathrm{s}^{-1}$, respectively. 
The results are in excellent agreement with their uncertainties, although they are derived from two independent measurements.
The uncertainty in \( f \) was estimated by propagating both statistical and systematic uncertainties associated with the ratio of the proportionality constant $r$ and with the estimation of the total volume within the hermetic detector, including its circulation system, $V_{\mathrm{I}}$. 
The volume $V_{\mathrm{I}}$ was determined by measuring the pressure and volume of gas introduced into the system, and was estimated to be $(1.8 \pm 0.2) \times 10^{-3}\,\mathrm{m^{3}}$. The uncertainty was estimated from the difference in $V_{\mathrm{I}}$ measured in two gas fills.
The steady-state ratio of radon concentrations at \( t \to \infty \) is correspondingly evaluated as 
\((1.1 \pm 0.1) \times 10^{-2}\) and \((1.1 \pm 0.2) \times 10^{-2}\) for the data obtained with the radon detector and the PMTs, respectively. 
The hermetic detector developed in this study reduced the radon concentration inside the detector to approximately 1\,\% of that measured outside.

\section{Discussion}
\label{sec:discussion}
In this section, we discuss the implications of the results presented in Sec.~\ref{sec:analysis} and estimate the expected performance when scaling the hermetic detector to the size required for the XLZD experiment, which is designed to employ TPC with an active mass of 60 to 80\,tonnes~\cite{XLZD:design}. 
Section~\ref{sec:discussion_scaling_and_leak} describes some key considerations for extrapolating the measurement results to XLZD, while Sec.~\ref{sec:calc_for_XLZD} provides a quantitative estimate of the resulting radon concentration in the scaled system. 
It should be noted that the hermetic detector in this study was not operated under cryogenic conditions.  
Therefore, any potential degradation of sealing performance due to differential thermal contraction of materials at operational temperature (approximately $-100\,^{\circ}$C) has not been accounted for.  
As described in Sec.~\ref{sec:circulation_system}, the gas-circulation system was designed with the future operation in LXe conditions in mind.  
As our next step, we plan to perform the same evaluation of radon-sealing performance under LXe conditions.

\subsection{Key Considerations for extrapolating the measurement results to XLZD}
\label{sec:discussion_scaling_and_leak}

The scaling factor to extrapolate the leakage flow measured in this experiment to XLZD is expected to depend on leakage paths. In the hermetic detector we developed, several possible radon leakage paths can be considered.
The first path is a leakage through the gasket, for which the total leak rate is expected to be proportional to its diameter.
Assuming a 60-tonne configuration, the TPC for XLZD would have approximate dimensions of 3\,m in both diameter and height. 
Given that the diameter of the gasket used in this study is approximately 6\,cm, the hermetic detector required for XLZD would be roughly 50 times larger in diameter. Therefore, the scaling factor is also expected to be 50 under this condition.
The second is the leakage through the piping connections linking the hermetic detector to the circulation system, whose contribution is expected to be proportional to the number of such connections. Since the number of piping connections in XLZD is not expected to increase significantly, the scaling factor should remain on the order of $\mathcal{O}(1)$ in this case.
Considering the dependence of the scaling factor on two distinct leakage pathways, we conservatively assume a scaling factor of 50 for the radon-leakage flow in XLZD.

In XLZD, radon is expected to emanate and exists in both the liquid and gaseous xenon phases outside the hermetic detector.
Since radon has a higher boiling point than xenon, it is known to accumulate in the liquid phase under static equilibrium conditions~\cite{xenon_radon}. 
However, in practice, the actual radon concentration depends on several conditions, such as the amount of emanation in each phase, the time constant for gas--liquid equilibration or the direction of the flow, and making it difficult to estimate the concentration in each phase precisely. 
Based on XENONnT studies of radon emanation~\cite{XENON:screening}, we conservatively assume that the number of radon atoms in the GXe is equal to that in the LXe for the following calculation. 
Since the diffusion coefficient of radon in LXe is roughly $1/500$ that in GXe~\cite{gasdiffusion,liquiddiffusion}, radon leakage is expected to be dominated by the transport in the gas phase.

\subsection{Estimation of radon concentration in XLZD experiment}
\label{sec:calc_for_XLZD}
In addition to the leakage discussed in Sec.~\ref{sec:discussion_scaling_and_leak}, the time evolution of the number of radon atoms inside the hermetic detector for XLZD also depends on the emanation of internal components and the removal process by the radon distillation system~\cite{xenon_radon}. 
Because the radon concentration inside the sealed volume is expected to be much lower than that outside, leakage from inside to outside can be neglected. 
The time evolution of the number of radon atoms inside the hermetic detector can thus be expressed as

\begin{equation}
\frac{dN_\mathrm{I}}{dt}
= -\frac{1}{\tau} N_\mathrm{I} 
+ \frac{\tau M f}{2 V_\mathrm{G-O}} K_\mathrm{O} 
+ K_\mathrm{I} 
- \frac{\rho_\mathrm{G} f_\mathrm{rr}}{W_\mathrm{I}} N_\mathrm{I},
\label{eq:radon_balance}
\end{equation}
where $N_{i}$, $W_{i}$, $V_{\mathrm{G-}i}$, $K_{i}$ and $\rho_\mathrm{G}$ are the number of radon atoms, the mass of xenon, the volume in the gas phase, the radon emanation rates, and the density of GXe, respectively.
For these parameters, the subscripts $i=\mathrm{I},\mathrm{O}$ denote the inside and outside of the hermetic detector, respectively.
$f_\mathrm{rr}$ is the xenon flow rate through the radon removal system, 
and $M$ is the scaling factor for the leakage flow $f$, which is set to 50 as discussed in Sec.~\ref{sec:discussion_scaling_and_leak}.
We set $W_\mathrm{I} = 60\,\mathrm{tonnes}$ from the XLZD design~\cite{XLZD:design} and $V_\mathrm{G-O} = 1\,\mathrm{m^3}$ based on the XENONnT cryogenic system~\cite{XENON:2023instrument_paper}.
The outer radon emanation rate in XLZD is taken to be $K_\mathrm{O} = 30\,\mathrm{mBq}$, following a previous study~\cite{freiburg_hermetic}. 

In Eq.~\ref{eq:radon_balance}, the first term on the right-hand side represents the decay of radon inside the hermetic detector, 
the second term corresponds to the radon inflow from the outer region through leakage, 
and the third and fourth terms account for the internal emanation and the removal by the distillation system, respectively. 

Using the measured leakage flow across the hermetic detector and the assumptions discussed in Sec.~\ref{sec:discussion_scaling_and_leak}, 
the radon leakage term can be calculated as
\begin{equation}
\frac{\tau M f}{2 V_\mathrm{G-O}} K_\mathrm{O} = 1.2 \times 10^{-2}\,\mathrm{[mBq].}
\end{equation}
This result indicates that implementing the hermetic-detector technique demonstrated in this study would enable a substantial reduction of radon-induced backgrounds in the XLZD detector. 
The leakage contribution is negligible compared to the internal emanation term of approximately 3\,mBq, 
as reported in a previous study~\cite{freiburg_hermetic}. 

Under this condition, the equilibrium radon concentration in the XLZD detector is determined primarily by the balance between the emanation inside the hermetic detector and the removal rate by the radon distillation system, which correspond to the third and fourth terms in Eq.~\ref{eq:radon_balance}, respectively.
The XLZD design goal of a radon concentration of $0.1\,\mu\mathrm{Bq/kg}$ can be achieved by assuming an emanation rate of 3\,mBq and a xenon mass of $60\,\mathrm{tonnes}$. 
Further reduction of radon concentrations may be achieved through studies on radon mitigation techniques, such as material selection~\cite{nEXO_material}, surface treatment~\cite{surface_coating}, and active removal systems such as cryogenic distillation~\cite{lowrad}.

\section{Conclusion and Outlook}
\label{sec:conclusion}

In this work, we developed and characterized a hermetic gaseous xenon detector employing a PTFE vessel sealed between two quartz flanges with mechanically compressed ePTFE gaskets.
The purpose of this study was to demonstrate an effective method for suppressing radon leakage into the sensitive detector volume, a key requirement for the direct detection of dark matter. 
Radon, especially \(^{222}\)Rn and its decay progeny, constitutes one of the dominant backgrounds in such experiments, and the target concentrations for next-generation detectors such as XLZD are an order of magnitude lower than those achieved in current experiments.

Our approach used a flange-based mechanical sealing system, employing narrow ePTFE gaskets and calibrated torque application to prevent radon leakage without relying on thermal contraction~\cite{freiburg_hermetic}. 
The design of the hermetic detector was experimentally optimized by comparing the baseline leak rates measured with a helium leak detector using natural helium present in ambient air.
The stability of the sealing performance was also investigated through long-term monitoring and a cooling test, demonstrating robustness against mechanical and thermal stresses.

We constructed a dedicated gas circulation and monitoring system capable of operating two fully isolated GXe loops. Using this system, we performed a long-duration radon injection experiment and tracked the radon concentrations inside and outside the hermetic detector using both electrostatic radon detectors and the PMTs. A detailed time-dependent analysis, based on first-order differential equations describing the interplay of decay and leakage flow, allowed us to extract the amount of leakage flow from the data. 
Leakage flows of \( (2.9 \pm 0.3) \times 10^{-11} \) and \( (2.6 \pm 0.4) \times 10^{-11} \)\,m$^{3}\,\mathrm{s}^{-1}$ were estimated using the data obtained with the radon detectors and PMTs. 
The corresponding suppression factors were calculated to be \( (1.1 \pm 0.1) \times 10^{-2} \) and \( (1.1 \pm 0.2) \times 10^{-2} \), respectively.

Based on the leakage flow estimated in this study, the expected radon sealing performance for XLZD was evaluated. 
Considering the scaling of the gasket length, the number of components (e.g., PMTs and cables), and the internal xenon volume, the estimated radon leakage is $1.2 \times 10^{-2}$\,mBq, which is negligible compared to the expected natural radon emanation inside the detector, typically 3\,mBq.

For future work, the following steps are planned. 
First, we will implement and test the hermetic detector under LXe conditions using a cryogenic system. 
This will enable us to quantify the impact of thermal contraction on sealing performance and to verify operational stability under realistic detector conditions. 
Second, we will introduce electrodes to operate the detector as a TPC. 
Design studies for single- and dual-phase TPC configurations are currently underway. 
Dedicated R\&D efforts, such as the development of coated microstrip electrodes on top of quartz flanges~\cite{amos_concepts}, are also in progress.

These developments are essential steps toward realizing a hermetic TPC for the XLZD experiment. 
Hermetically sealing the xenon inside the TPC with low-radioactivity components enables a significant reduction of radon-induced backgrounds, without relying solely on purification or material screening. 
The results of this study provide a strong foundation for adopting hermetic designs in future low-background noble-liquid detectors, not only for dark matter searches but also for other rare-event experiments where ultra-low background levels are essential.

\section*{Acknowledgment}
We thank the XMASS Collaboration and Y.~Takeuchi for providing the 2-inch PMT and the radon detector used in this study, respectively. We also thank M.~Yamashita for providing equipment for vacuum and GXe circulation during the initial study. This work was supported by JSPS KAKENHI Grant Numbers 19H05805, 20H01931, 21H04466, 24H00223, 24H02240, and JST FOREST Program Grant Number JPMJFR212Q.

% can use a bibliography generated by BibTeX as a .bbl file
% BibTeX documentation can be easily obtained at:
% http://www.ctan.org/tex-archive/biblio/bibtex/contrib/doc/

\bibliographystyle{ptephy}
\bibliography{reference}

\vspace{0.2cm}
\noindent
\let\doi\relax

\end{document}